\documentclass[]{aastex6}

\usepackage{amsmath}
\usepackage{multirow}

\begin{document}

\title{Circinus Galaxy Revisted with 10 Years of {\it Fermi}-LAT Data}

\author{Xiao-Lei Guo$^{1,2}$, Yu-Liang Xin$^{1}$, Neng-Hui Liao$^{3,1}$, Yi-Zhong Fan$^{1,2}$}

\affil{$^1$Key laboratory of Dark Matter and Space Astronomy, Purple Mountain Observatory, Chinese Academy of Sciences, Nanjing 210033, China}
\affil{$^2$School of Astronomy and Space Science, University of Science and Technology of China, Hefei 230026, Anhui, China}
\affil{$^3$Department of Physics and Astronomy, College of Physics, Guizhou University, Guiyang 550025,
Peoples Republic of China\\
nhliao@gzu.edu.cn (NHL); yzfan@pmo.ac.cn (YZF)}

\begin{abstract}
Circinus galaxy is a nearby composite starburst/AGN system. In this work we re-analyze the GeV emission from Circinus with 10 years of {\it Fermi}-LAT Pass 8 data. In the energy range of 1-500 GeV, the spectrum can be well fitted by a power-law model with a photon index of $\Gamma$ = $2.20\pm0.14$, and its photon flux is $(5.90\pm1.04) \times 10^{-10}$ photons cm$^{-2}$ s$^{-1}$. Our 0.1-500 GeV flux is several times lower than that reported in the previous literature,
which is roughly in compliance with the empirical relation for star-forming and local group galaxies and might be reproduced by the interaction between cosmic rays and the interstellar medium. The ratio between the $\gamma$-ray luminosity and the total infrared luminosity is near the proton calorimetric limit, indicating that Circinus may be a proton calorimeter. However, marginal evidence for variability of the $\gamma$-ray emission is found in the timing analysis, which may indicate the activity of AGN jet. More {\it Fermi}-LAT data and future observation of CTA are required to fully reveal the origin of its $\gamma$-ray emission.
\end{abstract}


\setlength{\parindent}{.25in}

\section{Introduction}
The $\gamma$-ray sky provides the crucial information of the Universe in the most extreme and violent forms. Since the Compton $\gamma$-ray Observatory era, blazars, an extreme subtype of Active Galactic Nuclei (AGNs; \citealt{UP95}), harbouring on-beam strong jets \citep{BR78}, are known as the dominant population of extragalactic $\gamma$-ray sources \citep{1999ApJS..123...79H}. Due to the relativistic effect, the jet emission is strongly amplified and highly variable \citep[e.g.,][]{1997ARA&A..35..445U}. The number of $\gamma$-ray blazars increases rapidly thanks to the successful performance of Large Area Telescope (LAT; \citealt{Atwood2009}) aboard {\it Fermi Gamma-ray Space Telescope}. Besides blazars, the AGNs with misaligned jets (MAGNs) have been found to be the $\gamma$-ray emitters as well \citep[e.g.,][]{2010ApJ...720..912A}. Moreover, the extended $\gamma$-ray emission associated with the radio lobes of nearby radio galaxies have been detected \citep{Abdo2010a,Ackermann2016}.

{\it Fermi}-LAT has also discovered the $\gamma$-ray emission from nearby star-forming galaxies, such as M82 and NGC 253 \citep{Abdo2010c}.
The interaction between cosmic rays (CRs) and the interstellar medium (ISM) can produce neutral pions, which subsequently decay into two $\gamma$-ray photons.
And the inverse-Compton scattering between CR electrons and interstellar radiation photons can also produce $\gamma$-ray emission.
Such paradigm applies to star-forming galaxies and the corresponding $\gamma$-ray emission may be detectable \citep[e.g.,][]{1989A&A...213L..12V,1991A&A...248..419A,1996ApJ...460..295P,2010ApJ...717....1L}.
Meanwhile, synchrotron radiation from the electrons moving in the interstellar magnetic fields contributes to the radio emission of star-forming galaxies \citep[e.g.,][]{1971A&A....15..110V,1973A&A....29..263V}.
With the remarkable linear empirical correlation between radio continuum (RC) and infrared (IR) emission luminosities \citep[e.g.,][]{1985A&A...147L...6D,1985ApJ...298L...7H,1992ARA&A..30..575C},
it is natural to have tight RC-$\gamma$ and IR-$\gamma$ luminosities correlations for the nearby star-forming galaxies \citep{Ackermann2012a}.
So far, there are four galaxies within the Local Group with $\gamma$-ray detections, including LMC \citep{2010A&A...512A...7A}, SMC \citep{2010A&A...523A..46A}, M31 \citep{2010A&A...523L...2A} and the Milky Way \citep{2012ApJ...750....3A}.
Several other star-forming galaxies have also been detected in the GeV band, including NGC 2146 \citep{Tang2014}, Arp 220
%\footnote{Note that there is a blazar candidate {\bf \color {red} CRATES J153456+233006} around Arp 220 which makes the association between   the $\gamma$-ray emission and Arp 220 ambiguous.}
\citep{Peng2016,Griffin2016}, and two hybrid systems with both AGN and star-forming activities, NGC 1068 and NGC 4945 \citep{Lenain2010}.
Recent attempts aiming to increase the number of such sources yield negative results \citep[e.g.,][]{Rojas-Bravo2016}.

Circinus is a nearby ($4.2\pm0.7$ Mpc; \citealt{Tully2009}) and almost edge-on ($\sim 65^\circ$; \citealt{Freeman1977}) spiral galaxy. It has been widely taken as a prototype Seyfert II AGN \citep[e.g.,][]{1994A&A...288..457O,1998A&A...329L..21O} and the nucleus is heavily obscured \citep[e.g.,][]{1999A&A...341L..39M,2004ApJ...614..135P}.
The circumnuclear starburst ring \citep[e.g.,][]{1994Msngr..78...20M,1998MNRAS.297...49E} makes Circinus a composite starburst/AGN system.
%In addition, large bipolar edge-brightened radio lobes consisting of central plume features that are likely jet-driven have been detected \citep{Elmouttie1998}.
In addition, the large edge-brightened radio lobes, likely caused by jet-driven outflow, are detected \citep{Elmouttie1998}.
The extended morphology has been resolved by {\it Chandra} in the X-ray band \citep{Mingo2012}. In the $\gamma$-ray regime, Circinus was unidentified in the initial phase of {\it Fermi}-LAT operation \citep{Nolan2012}. %since it is very close to the Galactic plane ($b = -3.8^\circ$).
It was claimed to be a $\gamma$-ray source by \cite{Hayashida2013} with 4 years of {\it Fermi}-LAT Pass 7 data. In fact, it is the only ``normal'' $\gamma$-ray Seyfert galaxy (i.e. not radio loud narrow line Seyfert Is; \citealt{2009ApJ...699..976A}) listed in the latest fourth {\it Fermi}-LAT AGN catalog \citep[4LAC;][]{4lac}. In comparison to other $\gamma$-ray star-forming galaxies, Circinus seems to well deviate from the RC-$\gamma$ and IR-$\gamma$ luminosities correlations. Moreover, the emission from radio lobes likely fails to give an acceptable description of the broadband spectral energy distribution \citep[SED;][]{Hayashida2013}. Motivated by its uniqueness and the mysterious origin of the $\gamma$-ray emission, in this work we revisit the $\gamma$-ray properties by analyzing 10 years of {\it{Fermi}}-LAT Pass 8 data.

The paper is organized as follows. In Section 2, we describe the data analysis routines and present our results. In Section 3 we discuss the physical origin of the GeV emission. Section 4 is a summary for this work.

\section{Fermi-LAT Data Analysis}

\subsection{Data Reduction}

We analyzed {\it Fermi}-LAT Pass 8 data in the region of Circinus galaxy recorded from August 4, 2008
to August 4, 2018 (Mission Elapsed Time 239557418-555033605). Considering that it locates in the
Galactic plane characterized by the complicated diffuse $\gamma$-ray emission, the events with energies
from 1 GeV to 500 GeV of ``Source''
type (evclass=128 \& evtype=3) were chosen. And the region of interest (ROI) is a
$10^\circ \times 10^\circ$ box centered at the core position of it
($\rm R.A. = 213.291^\circ, \rm Decl. = -65.339^\circ$).
We excluded the events with zenith angle greater than $90^\circ$ to reduce the contamination from the Earth Limb.
Furthermore, the data were filtered with {\tt (DATA\_QUAL>0)\&\&(LAT\_CONFIG==1)} as recommended to exclude some low quality events.
The data were binned into 27 logarithmic energy bins
and 100 $\times$ 100 spatial bins with a size of 0.1$^\circ$.
We analyzed the data using {\it Fermitools}
\footnote{http://fermi.gsfc.nasa.gov/ssc/data/analysis/software/} and the
instrumental response function of P8R3{\_}SOURCE{\_}V2.
The diffuse Galactic and extragalactic $\gamma$-ray backgrounds were modeled by {\tt gll\_iem\_v07.fits} and
{\tt iso\_P8R3\_SOURCE\_V2\_v1.txt}\footnote{http://fermi.gsfc.nasa.gov/ssc/data/access/lat/BackgroundModels.html}, respectively.
%The latest {\it Fermi}-LAT fourth source catalog \citep[4FGL;][]{fermi2019} was used to address all nearby sources.
The model also takes into account all nearby sources included in the latest {\it Fermi}-LAT fourth source catalog \citep[4FGL;][]{fermi2019}.
Circinus is associated with the point source 4FGL J1413.1-6519, whose spectrum is a single power-law.
During the fitting procedure, the normalizations and spectral parameters of 4FGL sources within the ROI, as well as the normalizations
of two diffuse backgrounds, were left free.
And the binned maximum likelihood analysis method with {\tt gtlike} was applied.

\subsection{Results}
\begin{figure*}[!htb]
%\centering
\includegraphics[width=3.5in]{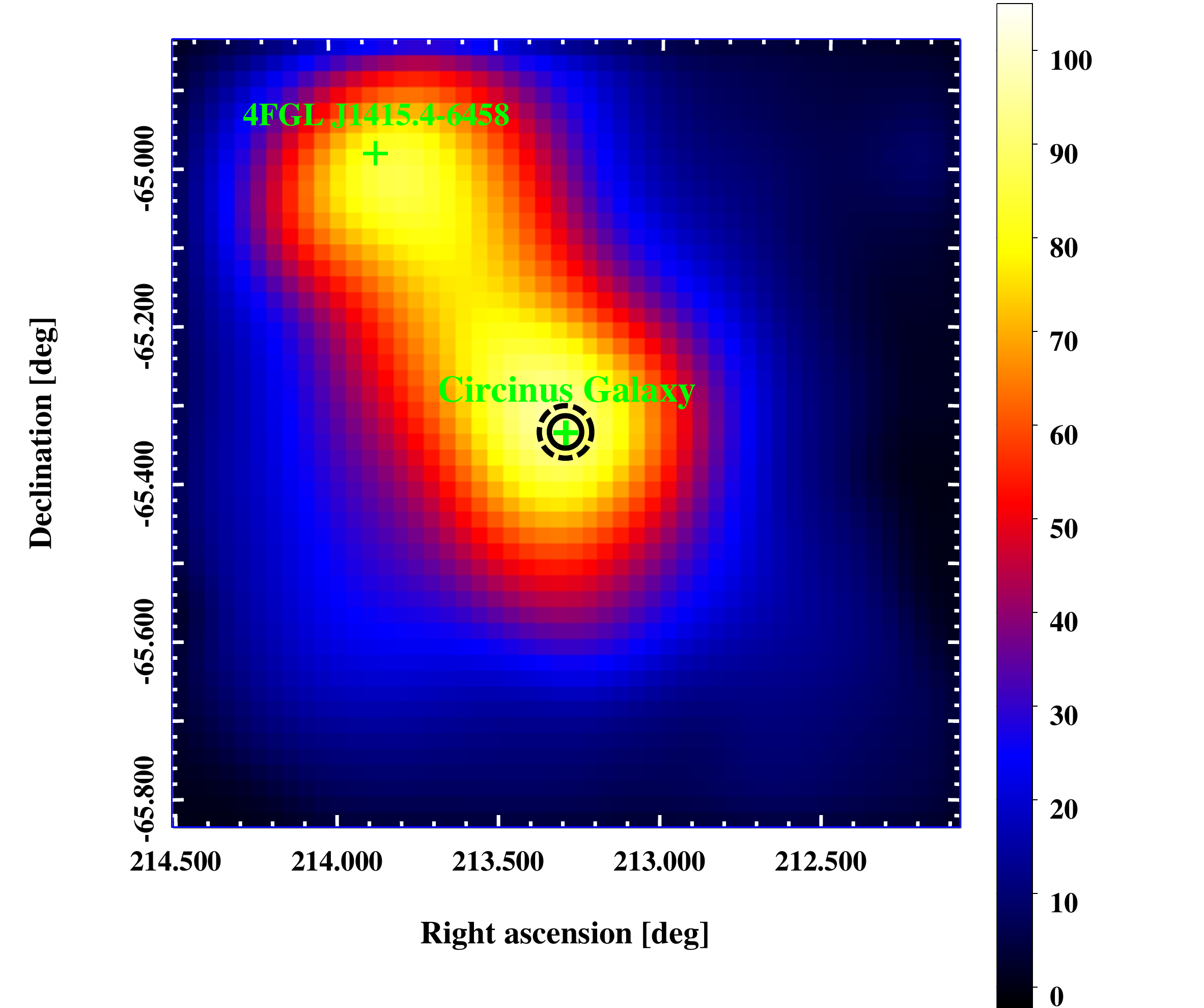}
\includegraphics[width=3.5in]{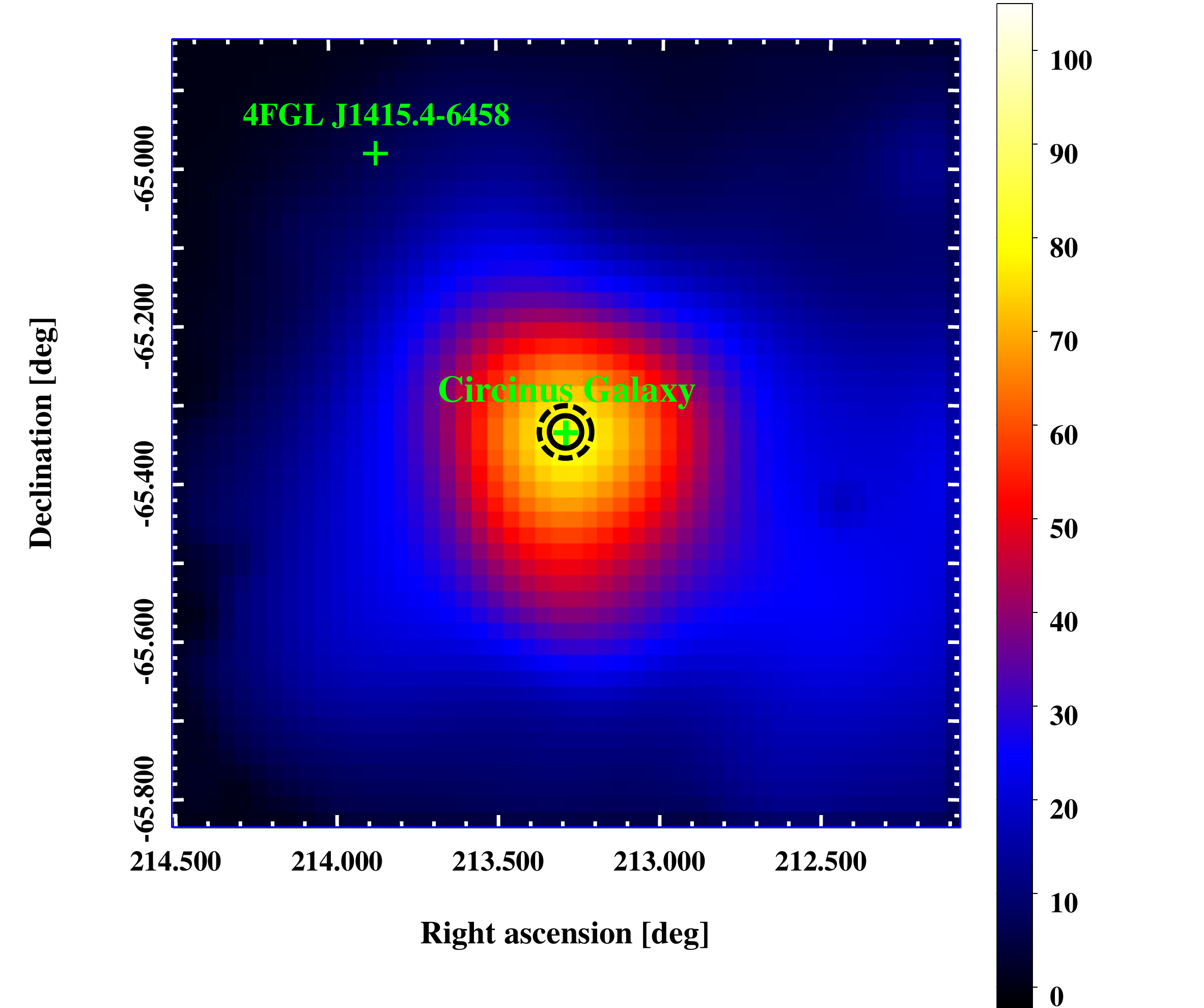}
\includegraphics[width=3.5in]{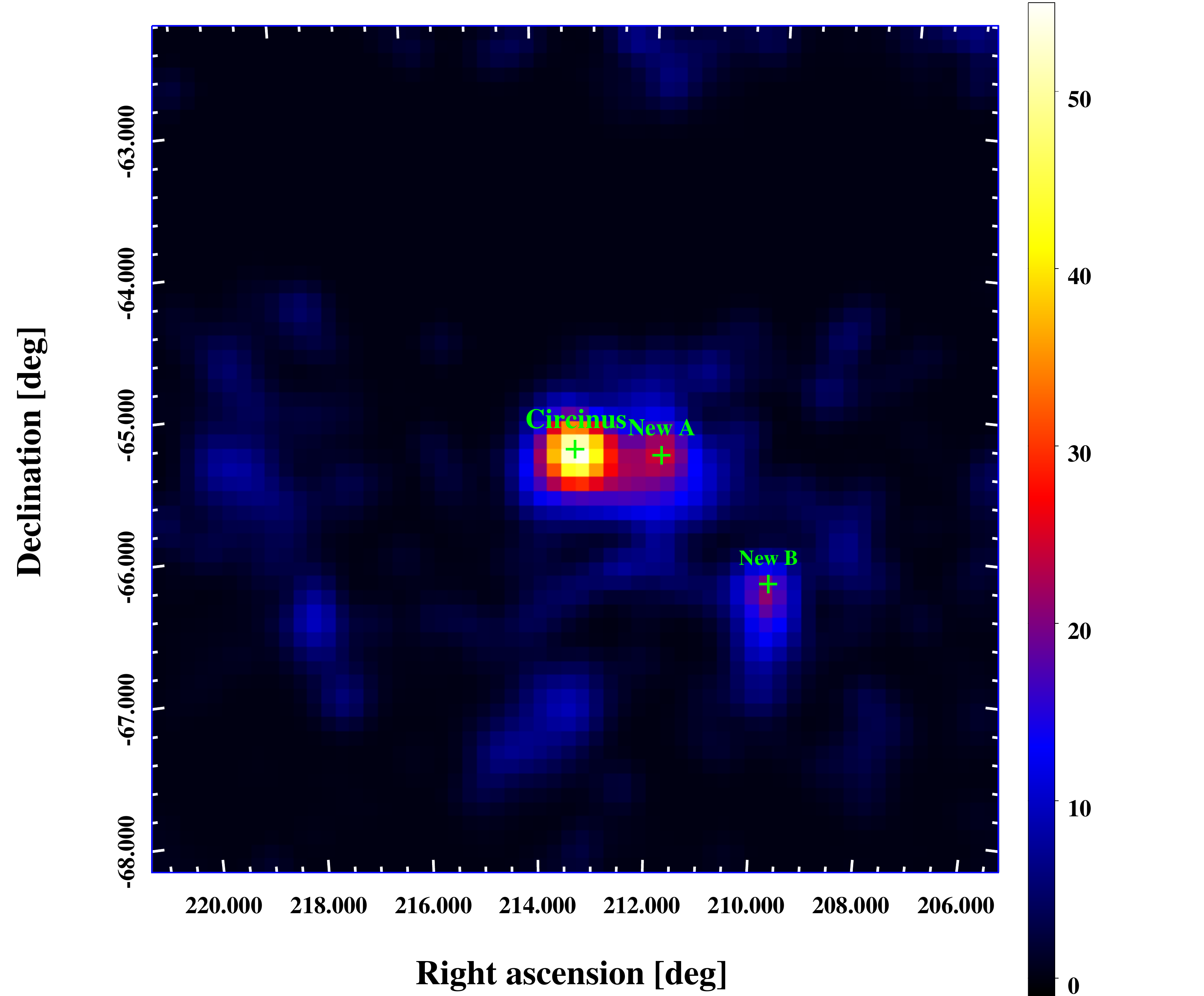}
\includegraphics[width=3.5in]{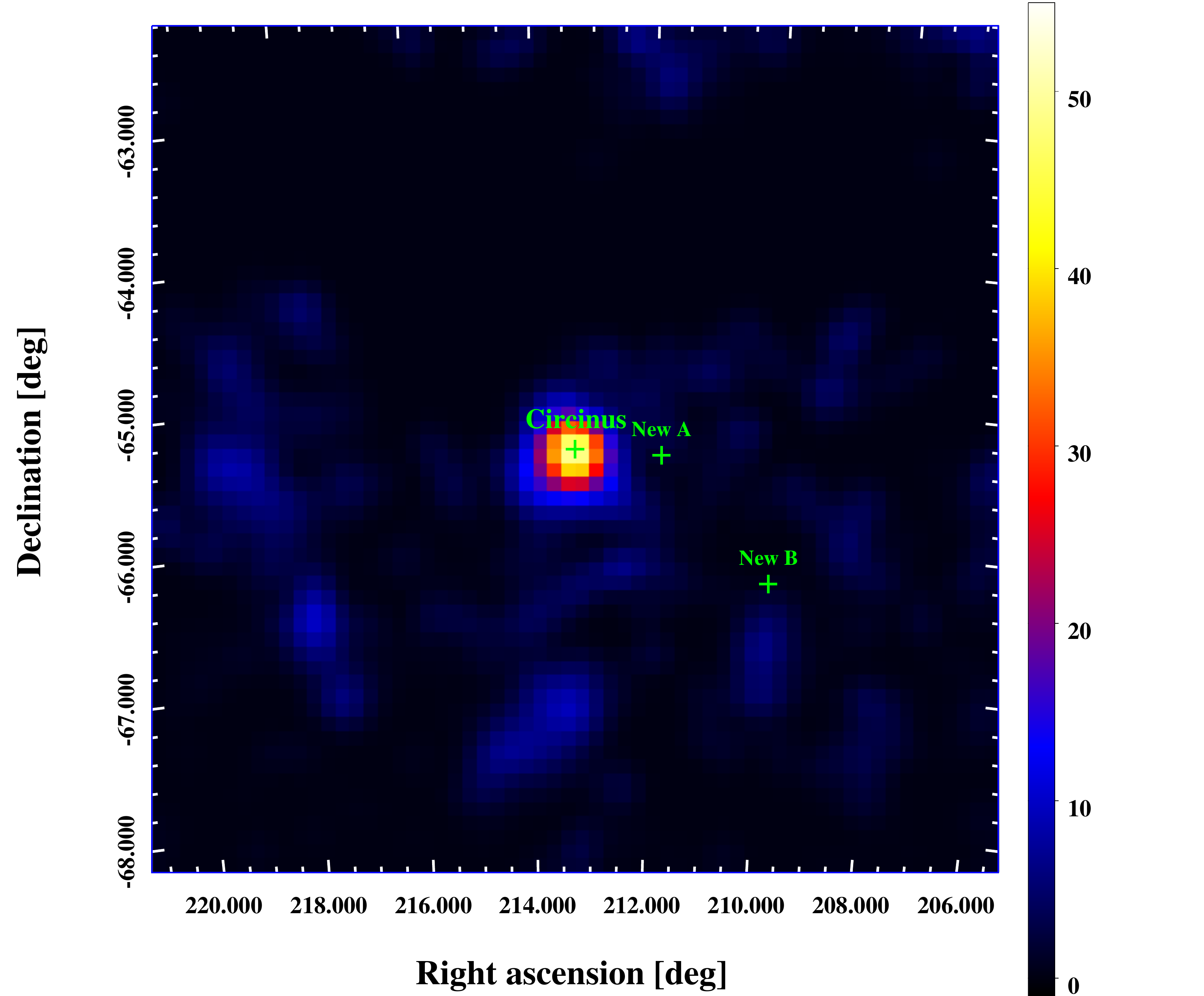}
%\vspace{5em}
\caption{Top panels: TS maps of $1^{\circ}\times1^{\circ}$ region centered on the core
position of Circinus galaxy with energies from 1 GeV to 500 GeV. The pixel size is
$0.02^\circ$ and the two TS maps are smoothed with a Gaussian kernel of
$\sigma$ = $0.04^\circ$. The left one is the TS map subtracting the
diffuse backgrounds and the 4FGL sources (except for 4FGL J1415.4-6458 and
Circinus). And the right one is the TS map with 4FGL J1415.4-6458
subtracted. The solid and dashed black circles represent the $1 \sigma$ and $2 \sigma$
error circles of the best-fit position of Circinus, respectively.
Bottom panels: $6^{\circ}\times6^{\circ}$ TS maps centered on Circinus with energies
above 1 GeV. Each pixel is $0.1^\circ$ and the TS maps are smoothed with a
Gaussian kernel of $\sigma$ = $0.2^\circ$. The left TS map is generated with model only
containing 4FGL sources and two diffuse backgrounds. And the right one is plotted with
model including New A and New B additionally.}
\label{fig:tsmap}
\end{figure*}

In our analysis, a point source, \textsc{4FGL J1415.4-6458}, locating only $0.429^\circ$
away from the core position of Circinus, was noticed (see the top panels of Figure \ref{fig:tsmap}).
This source was absent in \citet{Hayashida2013} possibly due to a significantly short observation time and the relatively limited quality of the Pass 7 data.
%and such a difference could be due to the more than twice as much as exposure of our data and the improvements of Pass 8 data relative to Pass 7 data.
After subtracting the emission from \textsc{4FGL J1415.4-6458}, there is still a significant
excess at the position of Circinus (see the upper right panel of Figure \ref{fig:tsmap}).
Moreover, we found some excesses except Circinus as shown in Figure \ref{fig:tsmap}, which were not included in the model.
We marked them as New A and New B, respectively.
%when a larger $6^{\circ}\times6^{\circ}$ TS map above 1 GeV (see the bottom panel of Figure \ref{fig:tsmap}.) was created.
Then the two new sources were added in the model as point sources with power-law spectra, and we re-performed the analysis.
The precise coordinates of Circinus, New A and New B are obtained by the {\tt gtfindsrc} command, which are listed in Table \ref{table:position}.
The best-fit position of Circinus is $\rm R.A. = 213.292^\circ, \rm Decl. = -65.3378^\circ$,
and its 1$\sigma$ error circle is $0.02^\circ$.
The angular distances between Circinus and New A/New B are $0.61^\circ$ and $1.67^\circ$, respectively.
Therefore, New A and New B have little influence on Circinus above 1 GeV.
However, in the low energy range, the contamination may be significant.
We searched for the possible counterparts of New A and New B using {\tt simbad} \footnote{http://simbad.cfa.harvard.edu},
but failed to identify plausible counterparts.

After adding New A and New B to the model, we have a TS value of 75 for Circinus galaxy,
corresponding to a significance level of $7.94\sigma$ with 4 degrees of freedom (dof).
Our TS value is only a little higher than that reported in \citet{Hayashida2013},
though the amount of our data set is 2.5 times of theirs.
The best-fit photon index above 1 GeV is $\Gamma$ = $2.20\pm0.14$, and the
integrated photon flux is $(5.90\pm1.04_{\text{stat}} \substack{+0.32 \\ -0.30} \ _{\text{sys}}) \times 10^{-10}$ photons cm$^{-2}$ s$^{-1}$.
%The systematic errors are due to the systematic uncertainty in the LAT effective area, which is
The systematic errors are caused by the uncertainties of the effective area of LAT, which are
estimated by the bracketing $A_{\text{eff}}$ method
as recommended \footnote{https://fermi.gsfc.nasa.gov/ssc/data/analysis/scitools/Aeff\_Systematics.html}.
%We note that our integrated photon flux is much lager than the
The photon indices of New A and New B are $\Gamma$ = $2.45\pm0.30$ and $\Gamma$ = $2.18\pm0.22$, and
the integrated photon fluxes are $(3.48\pm1.05_{\text{stat}} \substack{+0.20 \\ -0.21} \ _{\text{sys}}) \times 10^{-10}$ photons cm$^{-2}$ s$^{-1}$
and $(2.99\pm0.85_{\text{stat}} \substack{+0.17 \\ -0.14} \ _{\text{sys}}) \times 10^{-10}$ photons cm$^{-2}$ s$^{-1}$, respectively.

We also carried out a global analysis using 10 years' {\it Fermi}-LAT data
above 100 MeV within a $20^\circ \times 20^\circ$ ROI, and New A and New B were included
in the model. The energy dispersion correction was taken into account.
The TS value of Circinus is 76 and the photon index is $\Gamma$ = $2.05 \pm 0.11$.
The photon flux of $(5.72 \pm 1.96_{\text{stat}} \substack{+0.21 \\ -0.17} \ _{\text{sys}}) \times 10^{-9}$ photons cm$^{-2}$ s$^{-1}$ and the
$\gamma$-ray luminosity of $L_{0.1-100 \; {\rm GeV}} = (1.17\pm0.44) \times 10^{40}$ erg s$^{-1}$ were obtained.
Our flux is lower than the value given for 4FGL J1413.1-6519 in 4FGL,
which is likely due to the identification of the two new point sources. And
the $\gamma$-ray luminosity is also less than that found by \citet{Hayashida2013}.
The TS values of New A and New B are 31 and 39, respectively.
To compare with the results of \citet{Hayashida2013} straightforwardly, we also analyzed the LAT data with the same time interval (2008 August 5 - 2012 August 5) and the same
energy range (0.1-100 GeV). Our analysis yields a TS value of 41 and a photon flux of $F_{0.1-100 \; {\rm GeV}} = (7.60 \pm 3.45_{\text{stat}}
\substack{+0.24 \\ -0.24} \ _{\text{sys}}) \times 10^{-9}$ photons cm$^{-2}$ s$^{-1}$,
and both of them are lower than those of \citet{Hayashida2013}.
The differences may be caused by the detection of the new point sources around Circinus, the improvement of the data set (e.g. improved event reconstruction, better energy measurements, and significantly increased effective area) or more precise diffuse
$\gamma$-ray backgrounds.

\begin{table}[!htb]
\centering
\caption {Best-fit positions of Circinus galaxy and the two new point sources in the energy band of 1-500 GeV}
\begin{tabular}{cccc}
\hline \hline
Source Name & R.A.  & Decl.   & TS  \\
\hline
Circinus                 & $213.292^\circ$ & $-65.3378^\circ$ & $75$ \\
New A                    & $211.820^\circ$ & $-65.3755^\circ$ & $24$ \\
New B                    & $209.888^\circ$ & $-66.2572^\circ$ & $32$ \\
\hline
\hline
\end{tabular}
\label{table:position}
\end{table}

\subsubsection{Variability}
\begin{figure}[!htb]
%\begin{figure*}[bhpt]
\centering
\includegraphics[width=3.5in]{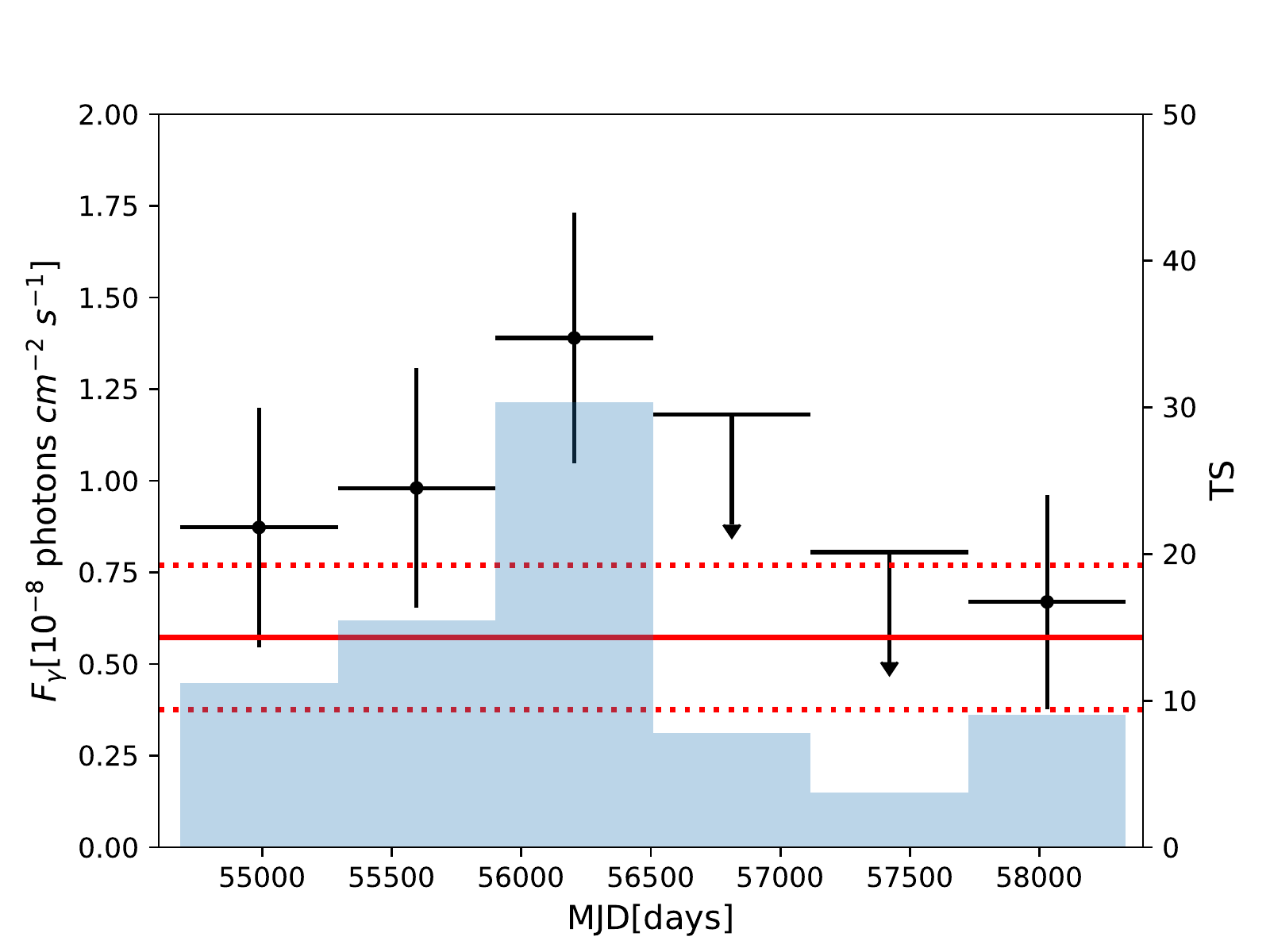}%
\hfill
\caption{The light curve of Circinus galaxy for 6 time bins with energy band from 100 MeV to 500 GeV. The red solid
horizontal line represents the average flux for the whole 10 years' data set, and
its $1\sigma$ statistic uncertainty is indicated by the red dotted lines. The blue histogram represents
the TS value for each time bin.}
\label{fig:lc}
\end{figure}

In view of the absence of obvious increase of the TS value compared to that in \citet{Hayashida2013},
% calculated with 10 years' data here
%when compared to that with 4 years data by \citet{Hayashida2013},
it is essential to explore the flux
variability of Circinus galaxy. The 10-year data with
energies between 100 MeV and 500 GeV were grouped into 6 time bins equally.
For the timing analysis, all
spectral indices of the sources included in the model were fixed to the
global fit values, and we repeated the likelihood analysis in each time bin.
For the time bin with a TS value smaller than 9,
an upper limit with 95\% confidence level was calculated. The light curve is shown in Figure \ref{fig:lc}.
We also calculated the variability index as defined in \citet{Nolan2012} and had TS$_{\rm var}$ = 12.8
corresponding to a significance level of 2.24$\sigma$, which is very marginal. Nevertheless,
we notice that the sum of TS values of the first five years is almost three times of that of the last five years.
%And the maximum ratio between flux in 6 time bins and the global average flux is $\sim$ 2.4.

\begin{figure*}[!htb]
%\centering
\includegraphics[width=3.5in]{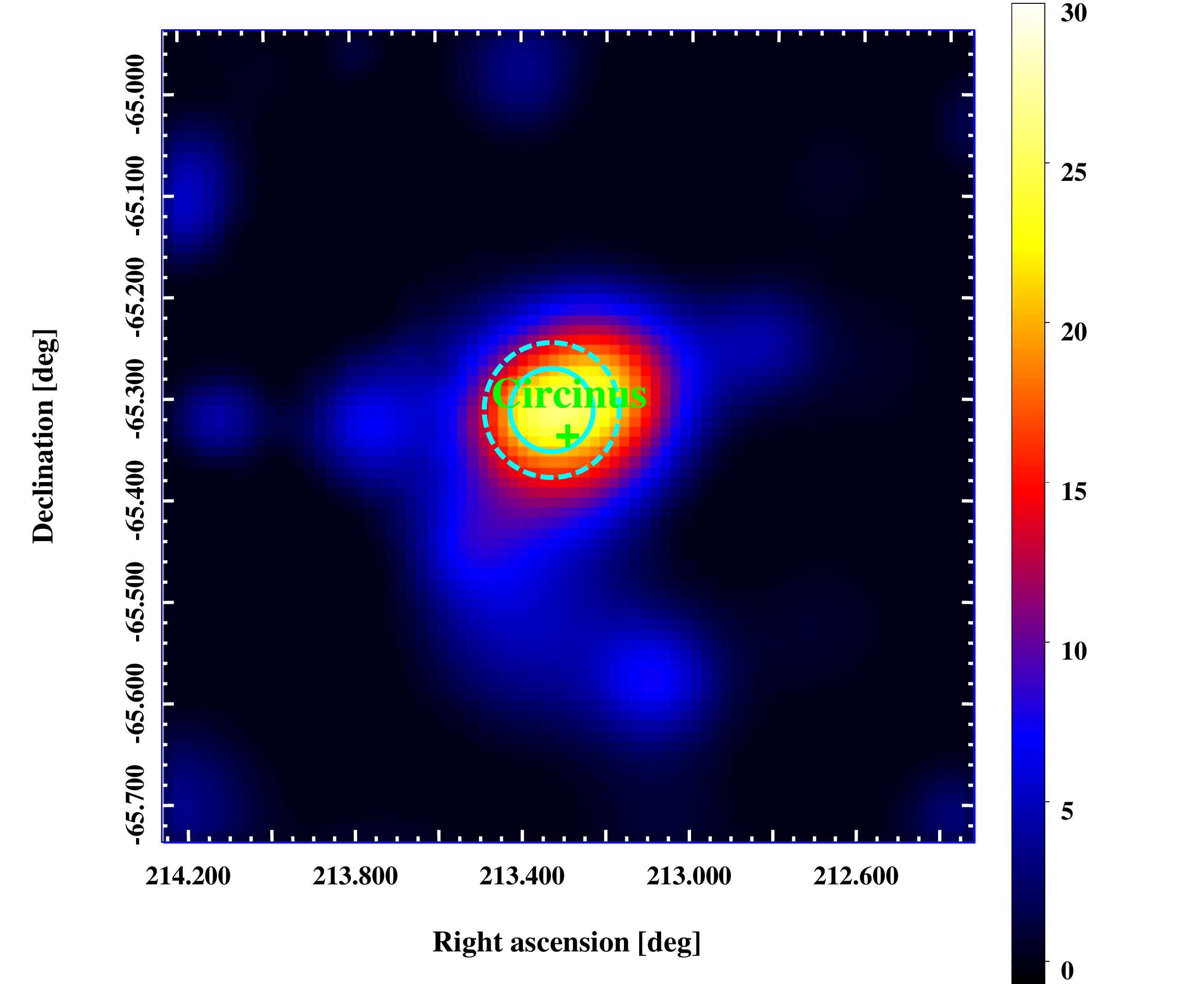}
\includegraphics[width=3.5in]{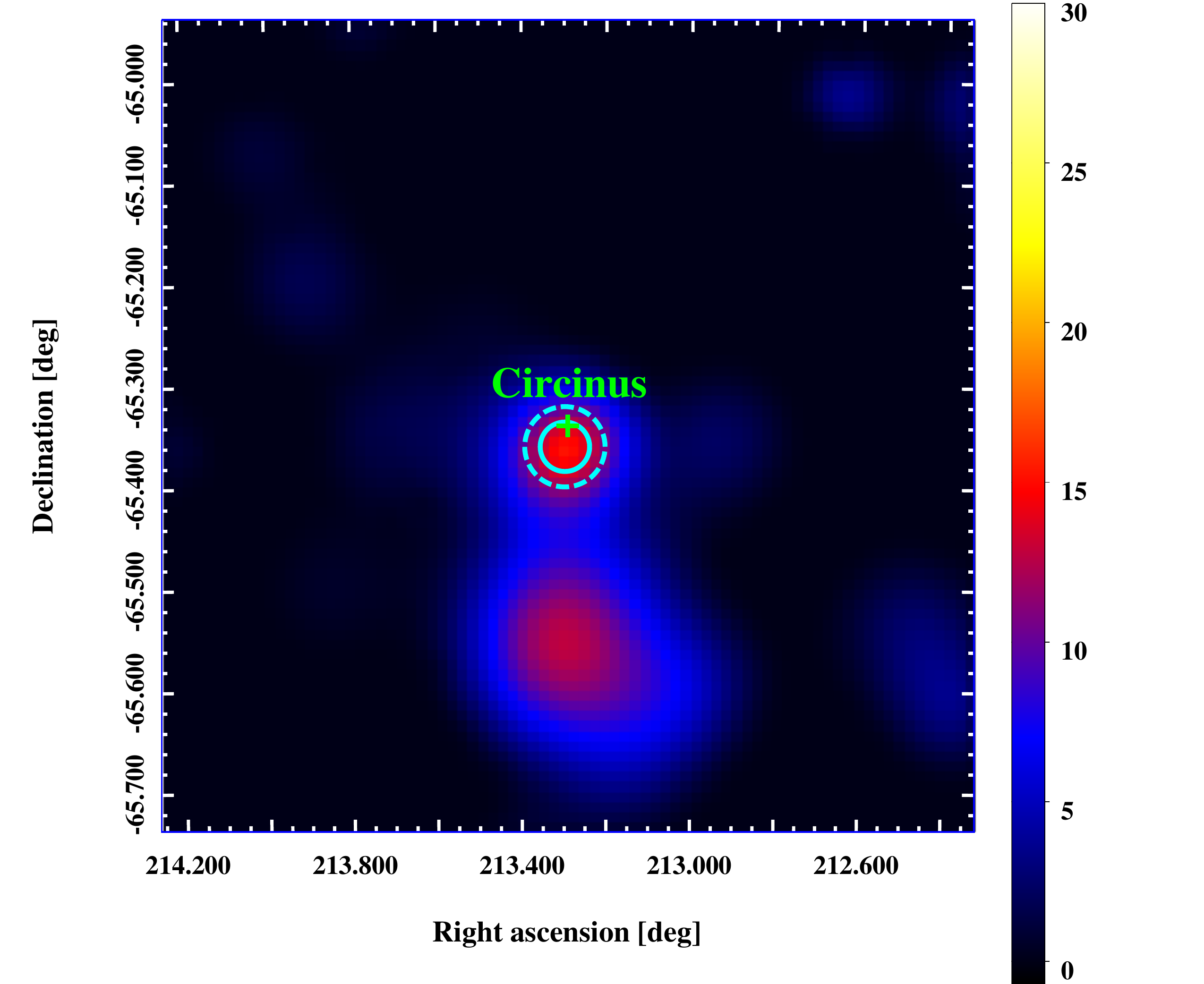}
%\vspace{5em}
\caption{$0.8^\circ \times 0.8^\circ$ TS maps for the data with energies above 10 GeV. Each pixel is 0.02$^\circ$.
The left one is for the first five years' data and the right one describes the result using the last five years' data.
The green plus is the core position of Circinus galaxy, and the solid and dashed cyan circles
present 1$\sigma$ and 2$\sigma$ error circles of the best-fit position.}
\label{fig:tsmap_10gev}
\end{figure*}

To further examine whether this source does have weak variability, the whole 10-year data above 100 MeV were divided into two equal parts by time, %two parts equally,
and maximum likelihood analyses were re-performed.
Since the Galactic diffuse $\gamma$-ray emission is stronger in the low energy band, we also analyzed the data with the energies above 1 GeV and 10 GeV in
the same time interval, respectively. Considering the low statistics in 10-500 GeV, unbinned maximum likelihood analysis method with a $5^\circ$ ROI
was adopted. The TS values and the photon fluxes for the first and last five years in three different energy bands are summarized in Table \ref{table:variability}.
%While events with different energy bands were analyzed,
In all scenarios,
the TS values of the first five years are about two times of that for the last five years,
which is evident in Figure \ref{fig:tsmap_10gev},
and the photon flux drops by a factor of $\sim 1.5-2$, suggesting a possible variability of the GeV emission from Circinus on timescales of years.
However, the evidence is still too weak to draw a firm conclusion.

\begin{table}[!htb]
\centering
\caption {TS values and photon fluxes of Circinus galaxy for the first and last five years}
\begin{tabular}{c|cc|cc}
\hline \hline
  Date      &\multicolumn{2}{c|}{August 4, 2008 - August 4, 2013} & \multicolumn{2}{c}{August 4, 2013 - August 4, 2018}    \\
\hline
 Energy Range       &  TS  & Photon Flux &  TS & Photon Flux  \\
\hline
0.1-500 GeV  & $56$ &  $(7.98 \pm 2.95) \times 10^{-9}$ & $28$ & $(3.51 \pm 1.96) \times 10^{-9}$  \\
1-500 GeV  & $62$ &  $(7.35 \pm 1.50) \times 10^{-10}$ & $36$ & $(5.09 \pm 1.48) \times 10^{-10}$  \\
10-500 GeV  & $24$ &  $(5.18 \pm 2.26) \times 10^{-11}$ & $12$ & $(2.37 \pm 1.37) \times 10^{-11}$  \\
\hline
\hline
\end{tabular}
\label{table:variability}
\tablecomments{The unit of the photon flux is photons cm$^{-2}$ s$^{-1}$, and only 1$\sigma$ statistic error is listed.}
\end{table}

\subsubsection{Spectral Analysis}
\begin{figure}[!htb]
\centering
\includegraphics[width=3.5in]{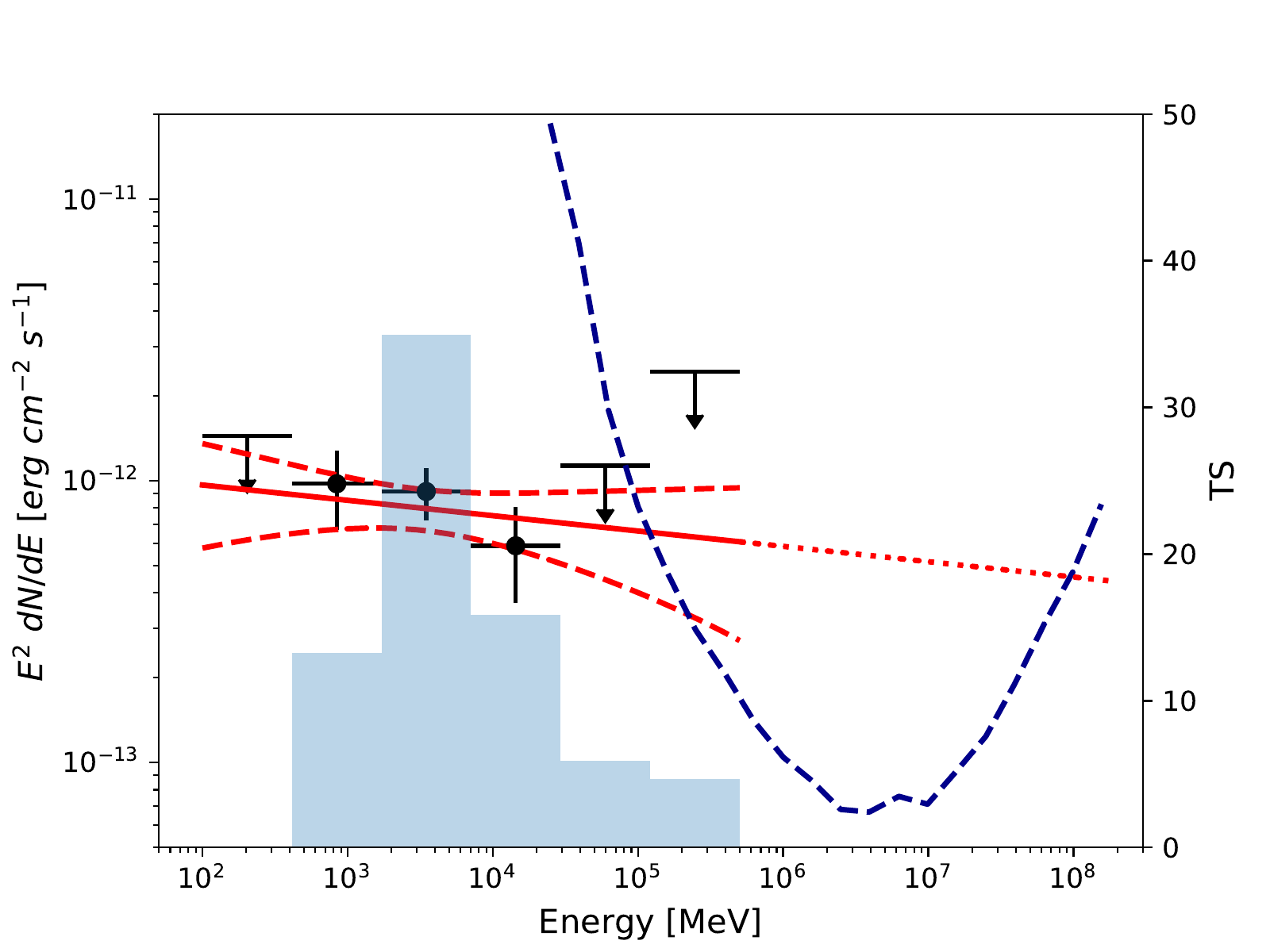}
\hfill
%\vspace{5em}
\caption{The $\gamma$-ray SED of Circinus galaxy. The global best-fit
power-law spectrum and its $1\sigma$ statistic error are plotted by the red
solid and dashed lines, respectively. The red dotted line represents the extrapolation
of the GeV spectrum. The TS values for the six energy bins are presented by
blue histograms. And the blue dashed line describes the differential energy flux sensitivities
of CTA-South with 50 hours' observation \citep{cta2017}.}
\label{fig:sed}
\end{figure}

To obtain the SED of Circinus, the data above 100 MeV
were binned into six log-equal energy bands. The likelihood analysis procedure is similar to
that in the timing analysis. And for energy bin with a TS value smaller than 9,
an upper limit at 95\% confidence level was given.
The result of {\it Fermi}-LAT SED is plotted in Figure \ref{fig:sed}.
%The flux for each bin is relatively smaller than that in \citet{Hayashida2013},
%which may mainly attribute to the removal of the contribution from the new nearby sources.
%So it is of great importance to consider the impact of these three sources during the whole analysis procedure.

\subsubsection{Spatial Extension and the Origin of $\gamma$-ray Emission}
To test the spatial extension of Circinus and explore the
origin of its $\gamma$-ray emission, 10 years of {\it Fermi}-LAT data with energies above 10 GeV within a $5^\circ$ ROI were selected,
considering the smaller PSF and weaker Galactic diffuse $\gamma$-ray background.
A point source and three uniform disks
with different radii were used as the spatial models.
The TS value of the point source model is 35.6 and the best-fit position is
R.A. = $213.302^\circ$, Decl. = $-65.3374^\circ$. The radii of the three uniform disks centered at the best-fit position of Circinus are
$0.1^\circ$, $0.15^\circ$ and $0.2^\circ$, respectively, and the corresponding TS values are
listed in Table \ref{table:template}. The larger the radius of the uniform disk is, the smaller the TS value is.
Therefore, a point source assumption is the best to describe the $\gamma$-ray emission.

\begin{figure}[!htb]
\centering
\includegraphics[width=3in]{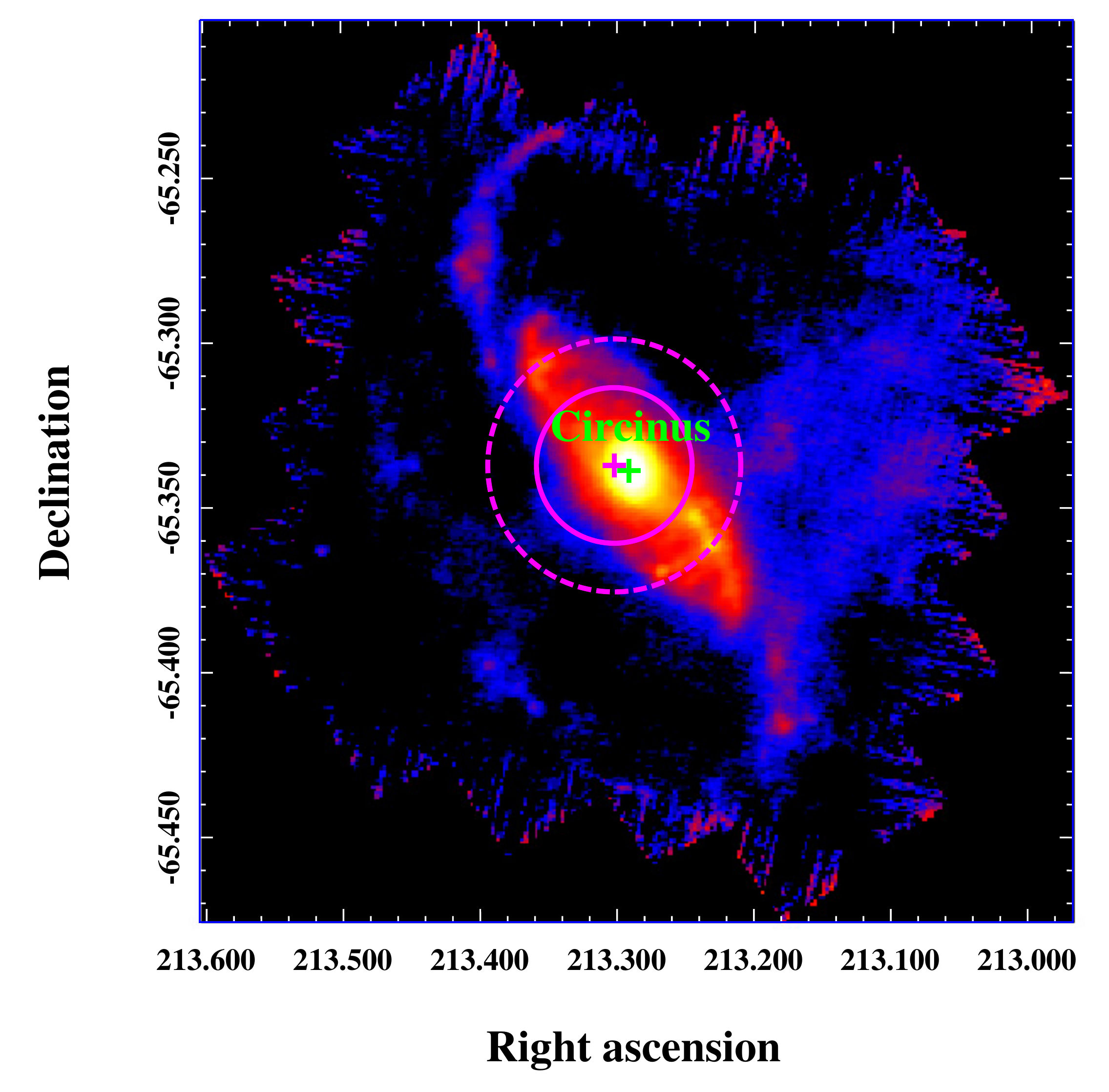}
\hfill
%\vspace{5em}
\caption{{\it Herschel}/PACS map at 160 $\mu$m. The magenta plus indicates the best-fit
position with data above 10 GeV. And the magenta solid and dashed circles
denote the 1$\sigma$ and 2$\sigma$ positional error circles, respectively.
The green plus is the core position of Circinus galaxy.}
\label{fig:ir}
\end{figure}

Since Circinus has large radio lobes \citep{Elmouttie1998} and the
emission from disk has been detected in the IR band,
%to probe where the $\gamma$-ray emissions of Circinus comes from,
the different geometrical spatial templates were also tested to probe the origin of its $\gamma$-ray emission.
We masked the core region of the {\it Herschel}/PACS map at 160 $\mu$m and used
it as the template to model the $\gamma$-ray emission from the disk. A point source located at
the west edge of the radio lobe (R.A. = $213.21^\circ$, Decl. = $-65.325^\circ$)
describes the lobe emission. And since our best-fit position is only $17.5''$ away
from the core region, we used a simple point source centered at the best-fit
position to model the core emission. The likelihood analysis were re-performed and the
calculated TS values for different spatial templates are reported in Table \ref{table:template}.
There are no significant differences among the different geometrical models,
and it is hard to pin down where the $\gamma$-ray emission comes from.
%The assumption that GeV emissions come from core region has the largest TS value.
%The TS values are close among the different geometrical models.
However, from the point of view of the best-fit position and its 1$\sigma$ error circle
shown in Figure \ref{fig:ir}, the core assumption may be favored.

\begin{table}[!htb]
\centering
\caption {TS values of Circinus galaxy with different spatial models}
\begin{tabular}{cccc}
\hline \hline
Spatial Model & TS Value & Degrees of Freedom \\
\hline
$0.1^\circ$ uniform disk                      & $28.0$ & $4$ \\
$0.15^\circ$ uniform disk                      & $22.8$ & $4$ \\
$0.2^\circ$ uniform disk                      & $20.2$ & $4$ \\
Point source (core assumption)                 & $35.6$ & $4$ \\
Point source (lobe assumption)                & $29.5$ & $4$  \\
{\it Herschel}/PACS map                    & $31.2$ & $2$  \\
\hline
\hline
\end{tabular}
\label{table:template}
\tablecomments{The 4 dof for the point source model include 2 spatial
and 2 spectral parameters.
For the {\it Herschel}/PACS map, only 2 dof of the spectral parameters
are considered.}
\end{table}

\section{Discussion}

\begin{figure}[!htb]
\centering
\includegraphics[angle=0,scale=0.35,width=0.5\textwidth,height=0.3\textheight]{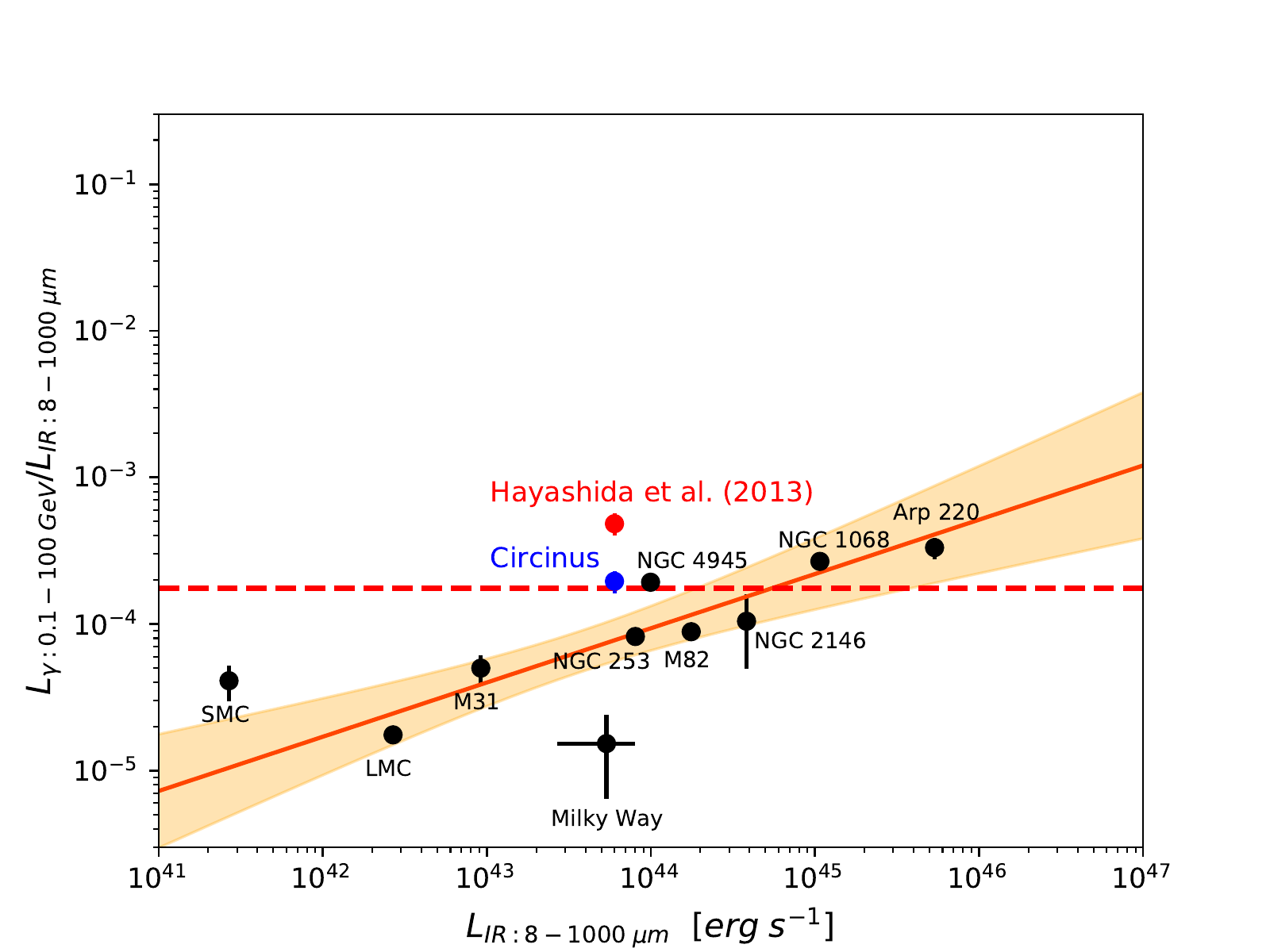}

\includegraphics[angle=0,scale=0.35,width=0.5\textwidth,height=0.3\textheight]{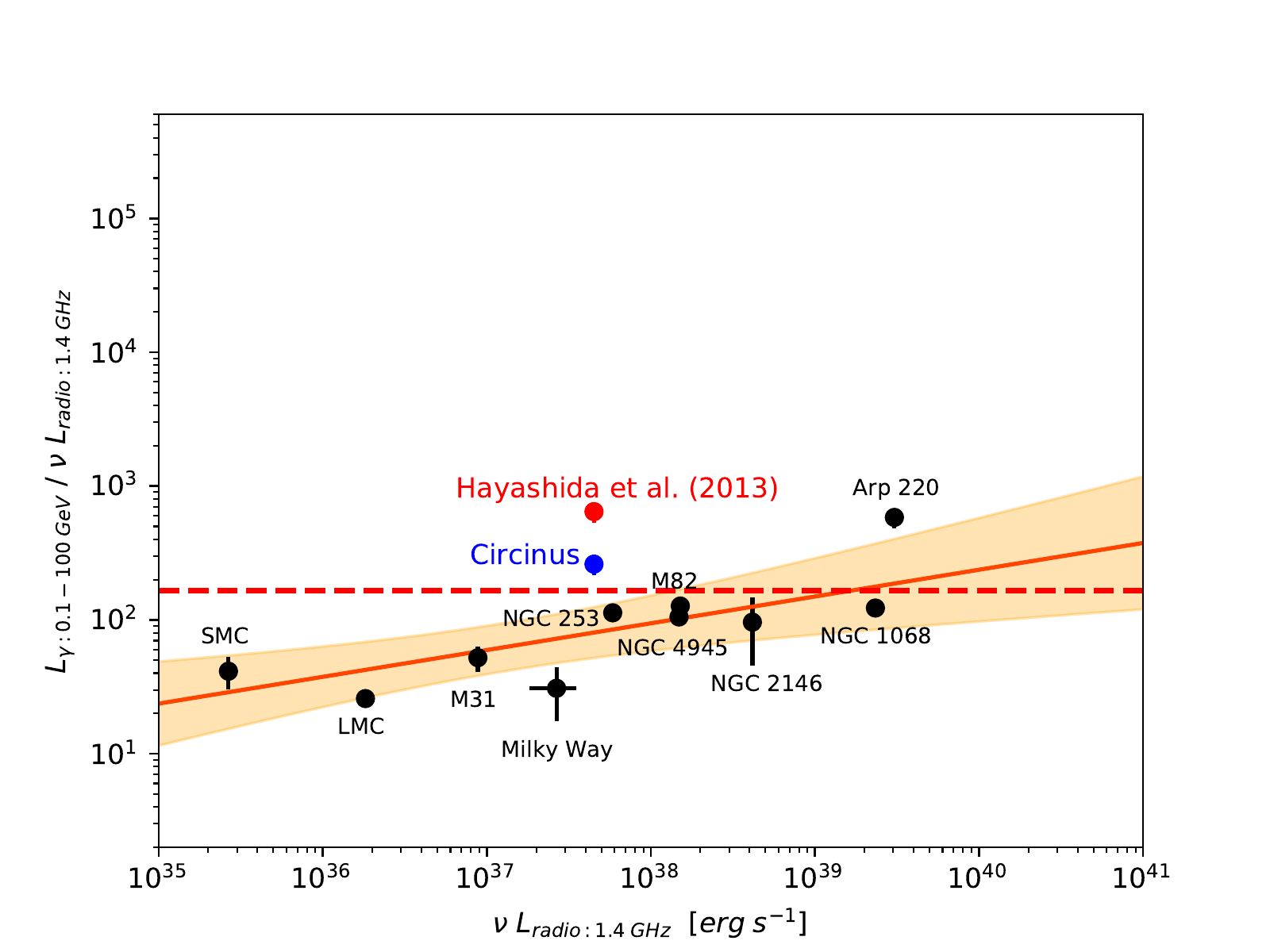}
\hfill
\caption{Top panel: the ratio between the $\gamma$-ray luminosity (0.1-100 GeV) and total IR
(8-1000 $\mu$m) luminosity. The best-fit relation
is plotted as the orange solid line and its 2$\sigma$ uncertainty
is shown as the shaded region. The red dashed line indicates the proton calorimetric limit.
And the red and blue dots describe the ratio of Circinus computed in \citet{Hayashida2013} and this work.
Bottom panel: the ratio between the $\gamma$-ray luminosity (0.1-100 GeV) and the radio continuum luminosity at 1.4 GHz.
The infrared data are taken from \citet{Gao2004}. The $\gamma$-ray data are taken from \citet{Ackermann2012a},
\citet{Tang2014}, \citet{Peng2016} and \citet{Wojaczynski2017}.
And the radio data are taken from \citet{Condon1990}, \citet{Wright1990} and \citet{Yun2001}.}
\label{fig:relation}
\end{figure}

It has been widely suggested that the $\gamma$-ray emission (above 0.1 GeV) from star-forming galaxy is
dominated by neutral pion decay resulting from the interaction between
CRs and ISM \citep[e.g.][]{Torres2004,Rephaeli2010,Lacki2011}. The CRs are primarily
accelerated by supernova remnants (SNRs), and the total CR injection power
is related to SN rate, the kinetic energy released by SN ($E_{\rm SN}$) and
the fraction of its kinetic energy transferred into CRs ($\eta$).
The SN rate can be assumed to be a constant fraction of star formation rate (SFR),
and the typical value of the kinetic energy released by SN is $10^{51}$ erg.
As suggested in \citet{Ackermann2012a}, the SFR can be well
traced by the total IR (8-1000 $\mu$m) luminosity that is an approximate calorimetric measure of
radiation from young stars, and \citet{Kennicutt1998} offered a conversion ratio that is given by

\begin{equation}
\frac{\rm SFR}{{\rm M}_{\odot}\:{\rm yr}^{-1}}=1.7 \epsilon \times 10^{-10} \frac{L_{8-1000\; \mu{\rm m}}}{L_{\odot}}.
\label{eq_sfr_ir}
\end{equation}

\noindent
The factor $\epsilon$ is a constant depending on the initial mass function (IMF).
%Here, the $\epsilon$ = 0.79 will be used following \citet{Ackermann2012a}.
In this work, we take $\epsilon$ = 0.79 following \citet{Ackermann2012a}.
%for \citet{Chabrier2003}.
In addition, the radio continuum luminosity at 1.4 GHz is an estimator of SFR \citep{Yun2001}, which can be expressed as

\begin{equation}
\frac{\rm SFR}{{\rm M}_{\odot}\:{\rm yr}^{-1}}=(5.9\pm1.8) \epsilon \times10^{-22} \frac{L_{1.4 \; {\rm GHz}}}{{\rm W\:Hz}^{-1}}.
\label{eq_sfr_rc}
\end{equation}

\noindent For a proton calorimeter, the CR protons would interact with ambient gas before
escaping from the galaxy. As usual, the CR spectrum with a power-law index of $\Gamma_p$ = 2.2 is assumed.
In the case of a calorimeter,
the $\gamma$-ray luminosity has a linear scale of the SFR, i.e.,

\begin{equation}
L_{0.1-100 \; {\rm GeV},\pi^{0}}\vert_{{\tau_{\rm
res}} \approx \tau_{\rm pp}} = 5 \times 10^{39} \rm{erg\;s}^{-1}
\left(\frac{\rm SFR}{{\rm M}_{\odot}{\rm yr}^{-1}}\right)
\left(\frac{\mathit{E}_{\rm SN}}{10^{51} \rm{erg}}\right)
\left(\frac{\eta}{0.1}\right).
\label{eq_luminosity_calorimeter}
\end{equation}

\noindent

With the total IR luminosity of $0.6\times10^{44}$ erg s$^{-1}$ \citep{Hayashida2013},
the SFR of Circinus galaxy can be estimated to be $\sim 2.1 M_{\odot}{\rm yr}^{-1}$. %using Equation (\ref{eq_sfr_ir}).
In the calorimetric limit, the $\gamma$-ray luminosity is expected to be about $1.05 \times 10^{40}$ erg s$^{-1}$,
%according to Equation (\ref{eq_luminosity_calorimeter}),
which is well consistent with the observation value of $(1.17\pm0.44) \times 10^{40}$ erg s$^{-1}$.
%Thus the star-forming process can reproduce the observed $\gamma$-ray emission, and Circinus is probably a proton calorimeter.
In addition, the $\gamma$-ray luminosities for star-forming galaxies have a large scatter of $10^{37}$ erg s$^{-1}$
$-$ $10^{42}$ erg s$^{-1}$, and the value of Circinus is within the luminosity range.
Moreover, there is no significant difference of $\gamma$-ray spectral indices between Circinus and other star-forming galaxies \citep{Ackermann2012a,Tang2014,Peng2016}.
Thus the star-forming process can reproduce the observed $\gamma$-ray emission, and Circinus is probably a proton calorimeter.

As mentioned above, there is a connection between SFR and $\gamma$-ray luminosities.
Actually, \citet{Thompson2007} predicted a linear relationship between far-IR and $\gamma$-ray for
the dense starbursts earlier.
With the detection of $\gamma$-ray emission from star-forming systems,
\citet{Ackermann2012a} established an empirical relation for star-forming
and local group galaxies between the $\gamma$-ray (0.1-100 GeV) and the
total IR (8-1000 $\mu$m) luminosity. \citet{Hayashida2013} found that the
$\gamma$-ray luminosity of Circinus was well above the scaling relation.
Since the $\gamma$-ray luminosity we calculated is lower than they found before, the discrepancy between
Circinus and the empirical relation is decreased.
Here, we update such relation with also NGC 2146 \citep{Tang2014} and Arp 220 \citep{Peng2016}.
%for comparison,
%we drawn a similar relation with NGC 2146 and Arp 220 which are not
%considered by \citet{Hayashida2013} (see Figure \ref{fig:relation}).
The scaling relation %between $\gamma$-ray luminosity and total IR luminosity
is fitted by a power-law,

\begin{equation}
\log \left(\frac{L_{0.1-100 \; {\rm GeV}}}{{\rm erg \; s}^{-1}}\right) =
\alpha \log \left(\frac{L_{8-1000 \; \mu{\rm m}}}{10^{10} L_\odot}\right)
+ \beta  .
\end{equation}

\noindent We also adopt a similar relation between $\gamma$-ray luminosity and radio
continuum luminosity at 1.4 GHz, i.e.,

\begin{equation}
\log \left(\frac{L_{0.1-100 \; {\rm GeV}}}{{\rm erg \; s}^{-1}}\right) =
\alpha \log \left(\frac{L_{1.4 \; \rm GHz}}{10^{21} \rm W Hz^{-1}}\right)
+ \beta .
\end{equation}

\noindent The best-fit parameters are summarized in Table \ref{table:relation}.

\begin{table}[!htb]
\centering
\caption {The best-fit parameters for the scaling relations}
\begin{tabular}{ccc}
\hline \hline
                & $\alpha$ & $\beta$ \\
\hline
$L_{0.1-100 \; {\rm GeV}}$-$L_{8-1000 \; {\rm \mu m}}$                 & $ 1.37 \pm 0.07$ & $39.40 \pm 0.07$ \\
 $L_{0.1-100 \; {\rm GeV}}$-$L_{1.4 \; {\rm GHz}}$               & $1.20 \pm 0.06$ & $38.95 \pm 0.09$  \\
\hline
\hline
\end{tabular}
\label{table:relation}
\end{table}
%\noindent where $\

In Figure \ref{fig:relation}, we present the ratio between $\gamma$-ray and total IR luminosities.
In the case of a proton calorimeter, the ratios among $\gamma$-ray, total IR and 1.4 GHz radio luminosities can be estimated by Equations (\ref{eq_sfr_ir}),
(\ref{eq_sfr_rc}) and (\ref{eq_luminosity_calorimeter}).
Circinus is basically in compliance with these relations, which is
different from \citet{Hayashida2013}.
Meanwhile, the fact that Circinus locates near the line of proton calorimetric limit
provides additional evidence as to be a possible proton calorimeter.

\begin{figure}[!htb]
%\begin{figure*}[bhpt]
\centering
\includegraphics[width=3.5in]{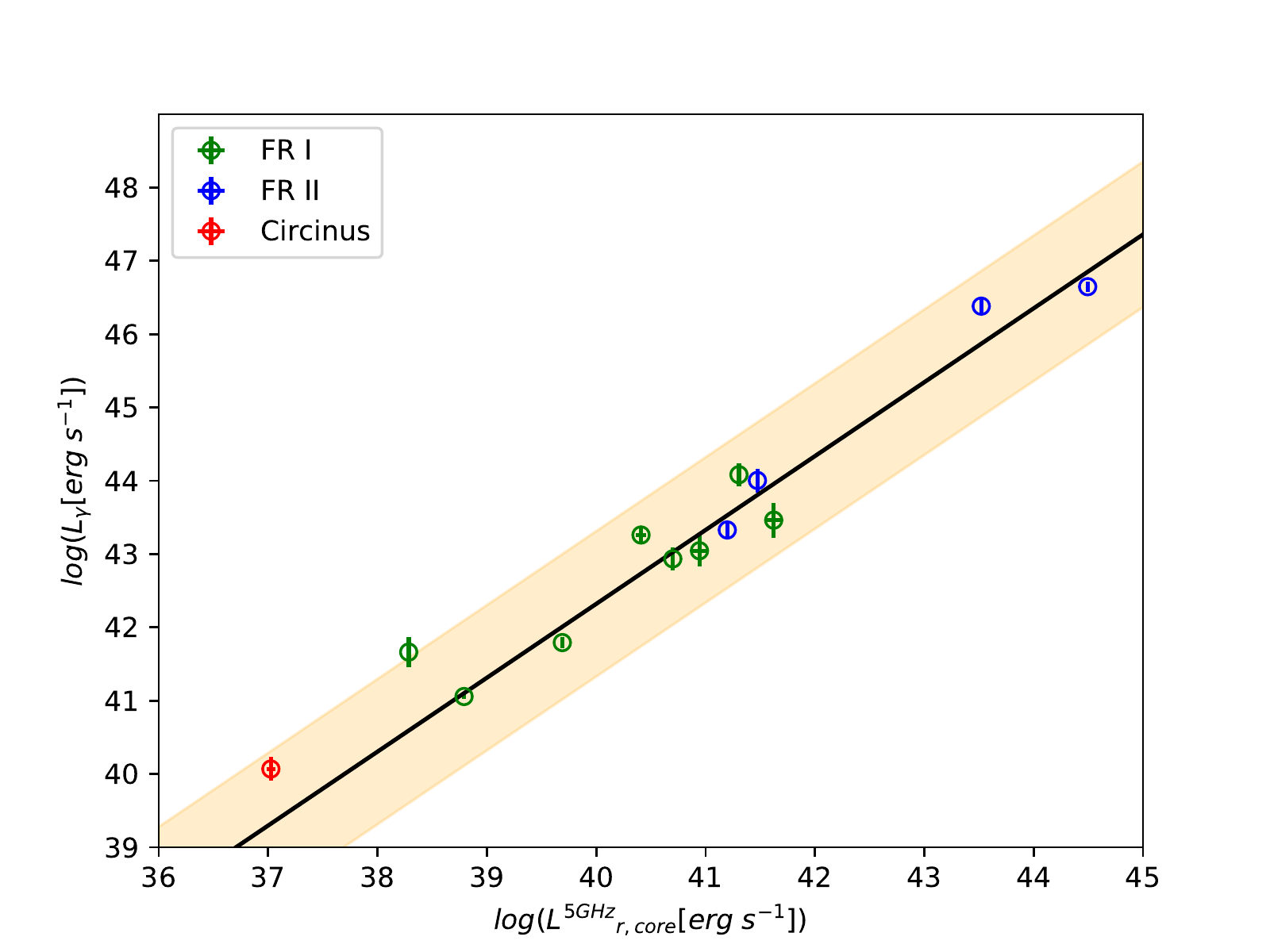}
\hfill
\caption{The $\gamma$-ray luminosity in the energy range of 0.1-100 GeV and radio-core luminosity at 5 GHz for the MAGNs and
Circinus. The $\gamma$-ray data for MAGNs are taken from \citet{Di Mauro2014}, and radio luminosity
at 5 GHz of Circinus is inferred from \citet{Elmouttie1998}. The black line and the shaded region are the best-fit
correlation and its 1$\sigma$ error presented in \citet{Di Mauro2014}, respectively.}
\label{fig:gamma-radio-core}
\end{figure}

Beyond the host galaxy, additional investigation of other possibilities about the origin of the $\gamma$-ray emission is worthwhile. In view of the strong and highly polarized edge-brightened radio lobe emission \citep{Elmouttie1998}, the $\gamma$-ray emission could be from the extended jet components which have been observed in nearby powerful radio galaxies \citep{Abdo2010a,Ackermann2016}. However, if the hint of the possible $\gamma$-ray variability on timescales of years is considered, such an explanation is likely not preferred. Otherwise, the faint arcsecond-scale core-jet may play an important role. Note that mild $\gamma$-ray variability on timescales from months to years has been occasionally detected in MAGNs (e.g., 3C 111; \citealt{Grandi2012}). Therefore, we look up the position of Circinus in the $L_{\rm radio-core}$ - $L_{\gamma}$ plot compared with known $\gamma$-ray MAGNs (see Figure \ref{fig:gamma-radio-core}). It is very interesting that Circinus is in compliance with the correlation of the MAGNs. If the contribution of the core-jet is indeed significant, Circinus could be a valuable target, shedding light on formation and energy dissipation of AGN jet in extreme environments. On the other hand, the $\gamma$-ray emission could be irrelevant to the AGN jet and generated through neutral pion production and decay in the two-temperature accretion flows around supermassive black holes (e.g., \citealt{2013MNRAS.432.1576N}). However, studies of searching $\gamma$-ray emissions from ``normal'' (non-jetted) Seyferts yield no detections of such sources (e.g., \citealt{2012ApJ...747..104A}).

%from relative steady star-forming process but from the AGN activity.
Similar to Circinus, NGC 1068 and NGC 4945 are also composite starburst/AGN systems. Although $\gamma$-ray variability has not been detected for NGC 1068 \citep{Lenain2010,Ackermann2012a}, its $\gamma$-ray emission is above the expectation of starburst activity \citep{Lenain2010,Eichmann2016}, suggesting the dominated emission may be from AGN. As to NGC 4945, a X-ray/$\gamma$-ray emission correlation suggests that the $\gamma$-ray emission is also dominated by AGN rather than the interaction between CRs and ISM \citep{Wojaczynski2017}. Though the ratios among $\gamma$-ray, total IR and radio luminosities of these three Seyfert galaxies are all close to or on the calorimetric limit line, it is less likely that all of them are proton calorimeters.

In the very high energy ($>$100 GeV) band, both starburst galaxies (NGC 253 and M82; \citealt{Acero2009,Abdalla2018,Acciari2009}), and nearby radio galaxies (e.g., M87 and Centaurus A; \citealt{2003A&A...403L...1A,2009ApJ...695L..40A}) have been detected. As the Figure \ref{fig:sed} shows, the extrapolation of the GeV spectrum of Circinus galaxy is above the differential sensitivity of Cherenkov Telescope Array \citep[CTA;][]{cta2017}. Therefore, Circinus could be detected by CTA in the future. Due to the limited angular resolution of {\it Fermi}-LAT, it is difficult to identify where the $\gamma$-ray emission comes from. However, the angular resolution of CTA, approaching 1 arc-minute at high energies \citep{cta2017}, is much better than {\it Fermi}-LAT. In the near future, CTA may pin down the origin of the $\gamma$-ray emission from Circinus.

\section{Summary}
In this work, we revisit the Circinus galaxy region using 10 years of Pass 8 {\it Fermi}-LAT data.
In the energy band from 1 GeV to 500 GeV,
Circinus is detected at a significance level of 7.9$\sigma$.
The spectrum can be well described by a power-law with a photon index of $\Gamma$ = $2.20 \pm 0.14$ and
the integrated photon flux is $(5.90\pm1.04_{\text{stat}} \substack{+0.32 \\ -0.30} \ _{\text{sys}}) \times 10^{-10}$ photons cm$^{-2}$ s$^{-1}$.
The $\gamma$-ray luminosity is $L_{0.1-100 \; {\rm GeV}} = (1.17\pm0.44) \times 10^{40}$ erg s$^{-1}$,
which is a few times lower than that found by \citet{Hayashida2013}.
As a result of the significant decrease of luminosity, the $\gamma$-ray emission is now roughly in compliance with the empirical
relation for star-forming and local group galaxies. The $\gamma$-ray emission from Circinus can be reproduced by the interaction
process between CRs and ISM. And it is possibly a proton calorimeter, considering the ratio between its
$\gamma$-ray and the total IR luminosities lies on the proton calorimetric limit.
However, the current data cannot rule out the presence of variability, indicating the $\gamma$-ray emission may be
dominated by the AGN jet activity instead of steady star-forming process.
To explore the origin of $\gamma$-ray emission and identify the variability of Circinus, more {\it Fermi}-LAT data and
the future observation of CTA are essential.

%\end{description}

\section*{Acknowledgments}
We acknowledge the anonymous referee for useful comments and suggestions,
the use of the {\it Fermi}-LAT data provided by the Fermi Science Support Center, the SIMBAD database
operated at CDS, Strasbourg, France and the data of Herschel provided by European-led Principal Investigator consortia and with important participation from NASA.
This work is supported by National Key Program for Research and
Development (2016YFA0400200), the National Natural Science
Foundation of China (Nos. 11433009, 11525313, 11722328, 11703093), and
the Joint Research Fund in Astronomy under cooperative agreement between
the National Natural Science Foundation of China and Chinese Academy of Sciences (grant No. U1738126).
%Natural Science Foundation of Jiangsu Province of China (No. BK20141444),
%and the 100 Talents program of Chinese Academy of Sciences.

\end{document}